\documentclass{aastex}

\shorttitle{Color Behavior of OJ 287} \shortauthors{Dai et al.}

\begin{document}
\title{Color Behavior Of BL Lacertae Object OJ 287 \\
During Optical Outburst}

\author{Yan Dai}
\affil{Department of Astronomy, Beijing Normal University, \\
Beijing 100875, China} \email{rdairye@mail.bnu.edu.cn}

\and

\author{Jianghua Wu}
\affil{Key Laboratory of Optical Astronomy, National Astronomical
Observatories, \\ Chinese Academy of Sciences, 20A Datun Road,
Beijing 100012, China}

\and

\author{Zong-Hong Zhu}
\affil{Department of Astronomy, Beijing Normal University, \\
Beijing 100875, China} \email{zhuzh@bnu.edu.cn}

\and

\author{Xu Zhou, Jun Ma}
\affil{Key Laboratory of Optical Astronomy, National Astronomical
Observatories, \\ Chinese Academy of Sciences, 20A Datun Road,
Beijing 100012, China}

\received{on 2010 August 25 by A.J.}

\accepted{on 2010 November 19 by A.J.}

\ccc{} \cpright{A.J.}{2010}

\begin{abstract}
This paper aims to study the color behavior of the BL Lac object OJ
287 during optical outburst. According to the revisit of the data
from the OJ-94 monitoring project and the analysis the data obtained
with the 60/90 cm Schmidt Telescope of NAOC, we found a
bluer-when-brighter chromatism in this object. The amplitude of
variation tends to decrease with the decrease of frequency. These
results are consistent with the shock-in-jet model. We made some
simulations and confirmed that both amplitude difference and time
delay between variations at different wavelengths can result in the
phenomenon of bluer-when-brighter. Our observations confirmed that
OJ 287 underwent a double-peaked outburst after about 12 years from
1996, which provides further evidence for the binary black hole
model in this object.

\end{abstract}


\section{Introduction}

Blazars, as a subset of active galactic nuclei (AGN), are
characterized by rapid and strong variability, and are those objects
with their relativistic jets pointed basically toward the observer
\citep{Urr95}. The jet is believed to originate from and be
accelerated by a rotating supermassive black hole surrounded by an
accretion disk. Blazars can be divided into two groups: BL Lac
objects and flat spectrum radio quasars (FSQR) The former ones show
no or very weak emission lines in their optical spectra, while the
latter ones show strong emission lines.

The BL Lac object OJ 287 is one of the best-observed blazars. It is
also the only blazar which shows convincing period of 12 years in
optical regime \citep{sil88}. In order to explain the 12-year
period, Sillanp\"{a}\"{a} et al. (1988) proposed a binary black hole
model for this object. Later observations found a double-peaked
structure for the periodic outburst and led other workers to
establish new models, but all based on a binary black hole system
\citep{leh96,sun97,kat97,vil98,val00,liu02,val08a,val08b,val09,val10,vil10}.

The color behavior of OJ 287 during outburst was investigated by
Sillanp\"{a}\"{a} et al. (1996). They found that the object showed a
stable $V$-$R$ color from 1994 to 1996. However, their color were
calculated by using the average magnitudes over one day. Since the
brightness of this object might change on intraday timescales, using
the average brightness will dilute the color change, especially that
on short timescales. In order to find out the genuine color, we
revisited the data from the OJ-94 monitoring project and used our
own data to figure out the relationship between the color and
magnitude of this object.

This paper is organized as follows. The revisit of OJ-94 data is
described in Section 2. Section 3 presents our monitoring and
results. Section 4 shows the simulations of the color behavior
resulted from different amplitudes and different paces of the
variations at different wavelengths. The conclusions are given in
Section 5.

\section{Revisit of OJ-94 Data}

In order to verify the predicted optical outburst of OJ 287 in late
1994 \citep{sil88}, an international, multi-waveband project was
launched. This is the OJ-94 project. It started in 1993 autumn and
ended in 1997. More than 50 workers from 10 countries were involved.
The monitoring was carried out at multiple wavelengths in the UV,
optical, infrared, and radio regimes. As a result, over 8,000 data
were collected during the monitoring period.

As mentioned in Section 1, the brightness of OJ 287 might change
during one day, as shown in Figure 1. Other cases of intraday
variability (IDV) in OJ 287 were reported by Carini et al. (1992).
When the magnitude of an object has little change in one day, it is
possible to use the average magnitude over that time to calculate
the color. But when the magnitude changes obviously on short
timescales, it is not a correct way to calculate the color using the
average magnitude. In order to get the genuine color, we reprocessed
the data of OJ-94 in two ways:

The first one was direct match. If the time interval of a $V$ and a
$R$ magnitude is less than five minutes, they were matched and a
color was computed. If one magnitude has multiple matches, the one
with the smallest time interval was used. The color-magnitude
diagram is given in Figure 2. In principle, the direct match is the
most reasonable way to calculate the color. However, this way might
be biased by the possible small systematic differences between
different telescopes involved in the OJ-94 project. Then we used  a
second way to calculate the color of OJ 287.

The second one was binning and match. The data were binned in half
an hour, and then the binned $V$ and $R$ magnitudes were used to
compute the color. The color-magnitude diagram is displayed in
Figure 3. The binning of the data helped to eliminate or reduce the
possible systematic differences between different telescopes. The
binning time of half an hour was chosen just because the variability
timescale of most blazars is longer than that time, except for a few
cases \citep[e.g.,][]{sas08,ran10}.

We used the linear least square method to calculate the
interdependency and got that the slope in Figure 2 is 0.02868, with
a correlation coefficient of 0.3022 and the significance level is
0.05. The slope in Figure 3 is 0.01818, with a correlation
coefficient of 0.2138 and the significance level is 0.05. For a
comparison, the slope of Sillanp\"{a}\"{a} et al. (1996) is just
0.00967 by converting their result from the slope of $V$-$R$
diagram. Therefore, we revisited the OJ-94 data and used two new
ways to calculate the color of OJ 287. The object was found to have
a bluer-when-brighter chromatism during its outburst in 1994-96.
This is different from and should be more reasonable than the result
drawn by Sillanp\"{a}\"{a} et al. (1996).

\section{Our Monitoring and Results}

\subsection{Observations and Data Reduction} \label{bozomath}

Our optical monitoring program of OJ 287 was performed with a 60/90
cm Schmidt telescope located at the Xinglong Station of the National
Astronomical Observatories of China (NAOC). Prior to 2006, there is
a Ford Aerospace 2048$\times$2048 CCD camera mounted at its main
focus. The CCD has a pixel size of 15 $\mu$m, and its field of view
is 58'$\times $58', resulting in a resolution of 1\farcs7
pixel$^{-1}$. At the beginning of 2006, the 2k CCD was replaced by a
new 4096$\times$4096 CCD. The field of view becomes 96'$\times$96',
resulting in a resolution of 1\farcs7 pixel$^{-1}$. The telescope is
equipped with a 15 color intermediate-band filters, covering a
wavelength range from 3000 to 10,000 {\AA}.

This paper includes data from 2005 January 29 to 2009 April 13.
Except for the nights with bad weather and those nights for other
targets, the actual number of nights with OJ 287 observations is
234. For the vast majority of nights, only one or two exposures were
made in each band. On only a few nights, more exposures were made
for IDV search. We used filters in $e$, $i$, and $m$ bands to
observe in 2005-06, and then changed to $c$, $R$, and $o$ bands from
the end of 2006 December. At the beginning of 2009, the $R$ band was
changed back to the $i$ band. The wavelengths of $c$, $e$, $i$, $m$,
and $o$ bands are 4210, 4920, 6660, 8020, and 9190 {\AA},
respectively. The $R$-band data has been published by Villforth et
al. (2010). The data at other wavelengths are reported here.

The data reduction procedures include positional calibration, bias
subtraction, flat-fielding, extraction of instrumental aperture
magnitude, and flux calibration. The average FWHM of stellar images
was about 3\farcs5 during our monitoring. So during the extraction,
the radius of the aperture was adopted as 3 pixels and the radii of
sky annulus were adopted as 7 and 10 pixels respectively. We used
the comparison stars 4, 10, and 11 in Fiorucci and Tosti (1996) for
the flux calibration of OJ 287. Their BATC $c$, $e$, $i$, $m$, and
$o$ magnitudes were obtained by observing them and three BATC
standard stars HD 19445, HD 84937, and BD +17 4708 on a photometric
night and are listed in Table 1. Then the magnitude of OJ 287 was
calibrated with respect to the average brightness of stars 4 and 10.
The star 11 acted as a check star. Its differential magnitudes were
calculated relative to the average brightness of star 4 and 10 for a
check of the accuracy of the observations. The observational log and
results are given in Tables 2 to 6. The columns are observation date
and time in universal time, Julian date, exposure time in second,
magnitude and error of OJ 287, and differential magnitude of star 11
(its nightly averages were set to be zero).

\subsection{Light Curve} \label{bozomath}

The light curves are shown in Figure 4. The data in the $R$ band are
also plotted here for a comparison. Generally, the light curves show
a double-peak outburst. The first outburst occurred in the period
from 2005 November 5 (JD 2,453,680) to 2005 December 24 (JD
2,453,729). It contains two sub-peaks. The second one was not well
identified by Villforth et al. (2010) due to their sparse time
coverage during that period. The second outburst, from 2007 October
13 (JD 2,454,387) to 2008 January 18 (JD 2,454,484), also contains
some sub-peaks, with the highest peak appeared on 2007 November 28
(JD 2,454,433). The time of the second outburst is consistent with
the predictions of different models \citep{sun97,val97,val00,liu02},
but the time of the first outburst has been advanced, which can be
explained as that the precession shifts the first outburst of each
outburst season progressively to earlier times relative to the mean
period \citep{val06}. Both of the outbursts contain some sub-peaks,
similar to the outburst in 1994-96, which can be explained by the
impact-tidal effects \citep{val09}, especially for the two sub-peaks
of the first outburst. The first sub-peak occurred because the
secondary black hole impact the accretion disk of the primary black
hole \citep{iav98}, while the second one was the tidal effect, i.e.,
tidally induced accretion flow would enhance the emission from the
jet.

From 2005 January 29 to 2006 November 19, the overall amplitude in
$e$, $i$, and $m$ band is 2.34, 2.24, and 2.14 mags, respectively.
And the overall amplitude in $c$, $R$, and $o$ band from 2006
November 26 to 2009 January 9 is 2.64, 2.59, and 2.55 mags,
respectively. The amplitude of variation tends to decrease with the
decrease of frequency. No IDV was found for OJ287 on the very
limited nights devoted to IDV search.

\subsection{Color Magnitude Diagram} \label{bozomath}

The color behavior of OJ 287 was studied based on our data. In
2005-06, we used $e$ and $m$ bands to calculate the relationship of
color and magnitude. Then $c$ and $o$ bands were used from the end
of 2006. The results are shown in Figures 5 and 6, respectively.

The linear least square method was used to calculate the correlation
coefficient. The slope we got in Figure 5 is 0.137, with a
correlation coefficient of 0.6689, and the significance level is
0.05. In Figure 6 the slope is 0.1164, with a correlation
coefficient of 0.626 and the significance level is 0.05. Figures 5
and 6 show strong chromatism of bluer-when-brighter.

The chromatism in Figures 5 and 6 are much stronger than Figures 1
and 2. The reason may be that the wavelength differences of the $c$
and $o$ bands (4980 {\AA}) and the $e$ and $m$ bands (3100 {\AA})
are much larger than that of the $V$ and $R$ bands (1070 {\AA}).

The same phenomenon of bluer-when-brighter was found for OJ 287 by
some authors. A strong correlation is observed by Brown et al.
(1989) between the near-infrared flux levels and the near-infrared
spectral slopes of OJ287, in the sense that the spectra are steeper
when the sources are fainter. Carini et al. (1992) found that there
does not appear to be a well-defined correlation between brightness
and color of OJ287; however, there is a general indication that when
the source is brighter, it tends to be bluer. According to the
comparison of the outburst of 1994 with those of 1971 and 1983,
Hagen-Thorn et al. (1998) got that there is a clear correlation
between the power of outburst and color indices of the variable
component in each event: the larger is the outburst, the bluer is
the source. Vagnetti et al. (2003) pointed out that eight BL Lac
objects including OJ287 tend to be bluer when brighter. Fiorucci et
al. (2004) reported that lots of BL Lac objects such as OJ 287
display the phenomenon of bluer when brighter. Wu et al. (2006) also
discovered a bluer-when-brighter chromatism in this object.
Villforth et al. (2010) studied the optical spectral index
variations and found a bluer-when-brighter trend. In fact,
bluer-when-brighter chromatism tends to be a general feature of BL
Lac objects \citep{vag03,rani10}.

\section{Simulation of Color Behavior}

Wu et al. (2007) mentioned that difference in amplitude and pace of
the variations at different wavelengths may both lead to color
change. Here we made some simulations to clarify it. We created two
light curves. The first one was made by the following steps. The
baseline is a sine curve with an amplitude of 2.0. A fluctuation
with a random amplitude less than 0.2 was added to the sine curve.
Then a random error less than 0.15 was added to each point. The
second light curve was produced by changing the variation amplitudes
or phases of the first light curve. Then we studied the color
behavior based on the two simulated light curves.

\begin{enumerate}
\setcounter{enumi}{0}
\item Effect of amplitude difference. The two light curves are different
in variation amplitude, with the second light curve having an
amplitude of 2.5, as shown in the left panel of Figure 7. The
color-magnitude diagram is displayed in the right panel of Figure 7.
Although the points traced a zig-zag route, the object evolves
basically along the diagonal path on the color-magnitude diagram. A
simple linear fit to the points gave a correlation coefficient of
0.9348, which indicates a strong bluer-when-brighter trend. If the
amplitude difference increases, the slope of the linear fit
increases, too.
\end{enumerate}

\begin{enumerate}
\setcounter{enumi}{1}
\item Effect of time delay. As shown in the left panel of Figure 8, the two
light curves are different in variation phase by 1/8 $\pi$. The
color-magnitude diagram is shown in the right panel. The numbers
indicate the time sequence. The object evolves on the
color-magnitude diagram along an elliptic orbit in the anticlockwise
direction. The major axis of the ellipse is along the line of
bluer-when-brighter. When we changed the phase difference from 1/8
$\pi$ to 1/4 $\pi$, the slope of the major axis of the ellipse
increases, and the ellipticity becomes smaller, but the
anticlockwise evolution remains. Except two extreme examples of
variability timescale of 15 minutes observed in S5 0716+714
\citep{sas08,ran10}, the variability timescales of optical bands are
longer than 1 hour on most occasions \citep{rom02,cel07}. However,
the observed time delay in optical band is usually shorter than 10
minutes \citep[e.g.,][]{qia00,vil00,pap03,sta06}. If $\pi$ (the
timescale), corresponds to one hour, then 1/4 $\pi$ equals 15 min,
so that it is no need to consider longer time delay in our
simulation. This loop path on the color-magnitude diagram has been
predicted theoretically by Kirk et al. (1998). Some observational
results at high energy were reported by several authors, as
mentioned by Wu et al. (2007).
\end{enumerate}

\begin{enumerate}
\setcounter{enumi}{2}
\item Effect of amplitude and time delay. The two light curves are different
in both amplitude and phase. The amplitudes of the two light curves
are 2.0 and 2.5, respectively, and the time delay is 1/8 $\pi$. Such
light curves are shown in the left panel of Figure 9. The right
panel gives the relationship of color and magnitude. The diagram
also shows as an ellipse in the anticlockwise direction. The major
axis of ellipse is also along the line of bluer-when-brighter, and
the ellipticity is larger than the one in Figure 8.
\end{enumerate}

The simulations confirmed that both amplitude differences and time
delay can result in the phenomenon of bluer-when-brighter. In all
three cases, the object tends to be bluer-when-brighter. The color
changes more significantly as the difference in amplitude or phase
increases. Moreover, the simulations provided a potential way to
distinguish between the factors that lead to the bluer-when-brighter
chromatism. If the object evolves along a diagonal path on the
color-magnitude diagram, the amplitude difference dominates the
color change. If there is a loop path on the color-magnitude
diagram, at least time delay is involved. Of course, the measurement
accuracy is a key factor in the identification of a diagonal or a
loop path on the color-magnitude diagram.

\section{Conclusions}

Based on the data from the OJ-94 monitoring project, we found that
OJ 287 showed a bluer-when-brighter chromatism in its outburst
period. Meanwhile, using the data obtained from the 60/90 cm Schmidt
Telescope of NAOC in 2005-09, we found an even stronger chromatism.
The amplitude of variation tends to decrease with the decrease of
frequency.

Our monitoring results proved that after the outburst in 1994 and
1996, the object underwent a new outburst in 2005 and 2007 after
about 12 years and showed a double-peaked structure. These results
further prove the binary black hole model of OJ 287. However, even
with these new observational data, there are still some inconsistent
results of the estimates for the masses of the two black holes
\citep[e.g.,][]{val08b,fan09}. The monitoring of this object till
next major outburst may help to resolve these conflicts.

We made some simulations and confirmed that both amplitude
differences and time delay can result in the phenomenon of
bluer-when-brighter. Moreover, the color measurement may be biased
by some technical and artificial effects \citep{wu10}. The relation
of the color behavior  to their central physics may be more complex
than expected before.

\begin{acknowledgements}
The authors thank the anonymous referee for constructive suggestions
and insightful comments. This work has made use of the data from the
OJ-94 project (http://www.astro.utu.fi/ research/oj94/) and has been
supported by Chinese National Natural Science Foundation grants
10633020, 10778714, 10873016. ZHZ acknowledges financial support
from the National Science Foundation of China under the
Distinguished Young Scholar Grant 10825313 and by the Ministry of
Science and Technology national basic science Program (Project 973)
under grant No. 2007CB815401.
\end{acknowledgements}

\begin{figure}
\includegraphics[width=8cm,height=8cm]{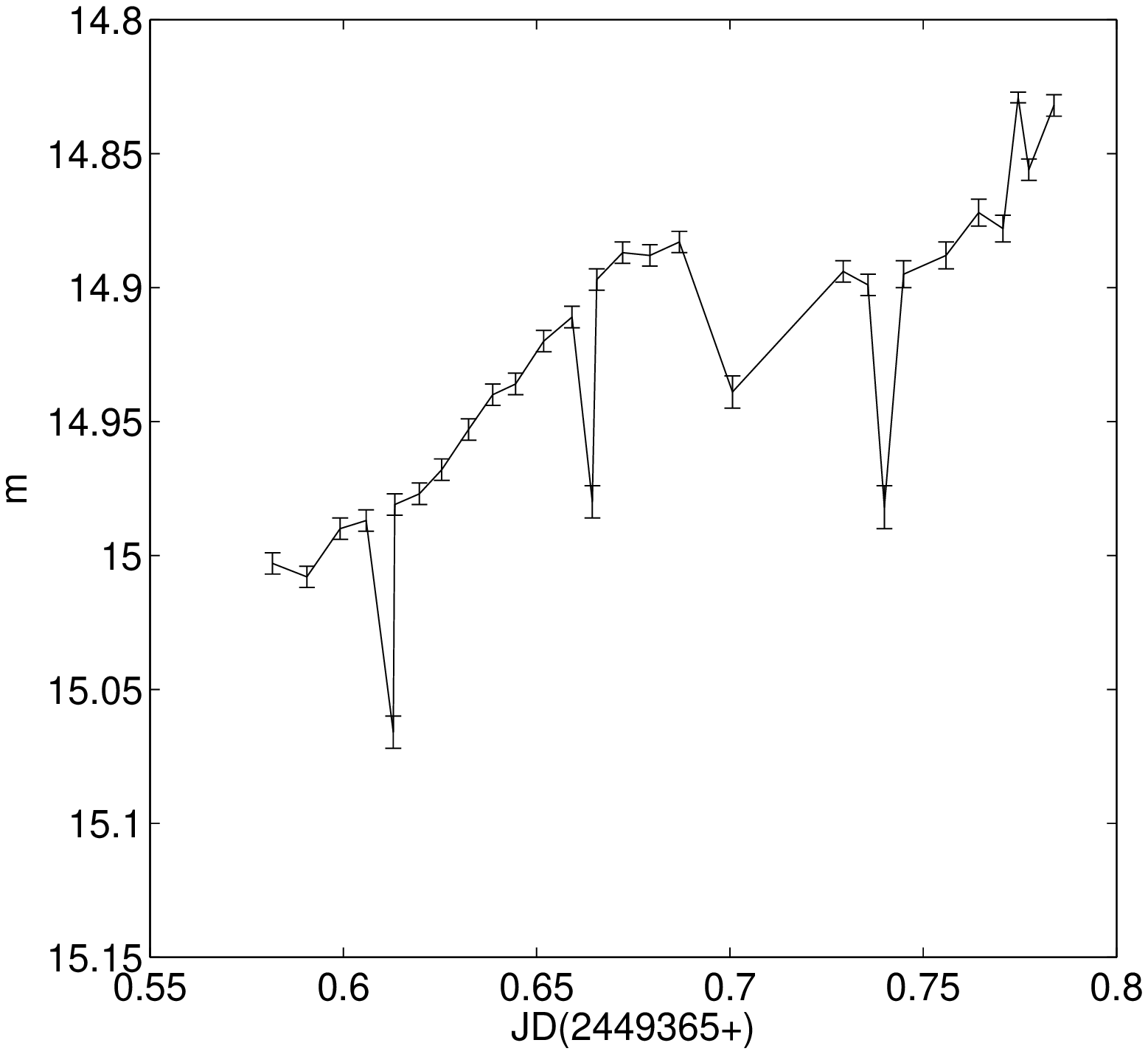}
\includegraphics[width=8cm,height=8cm]{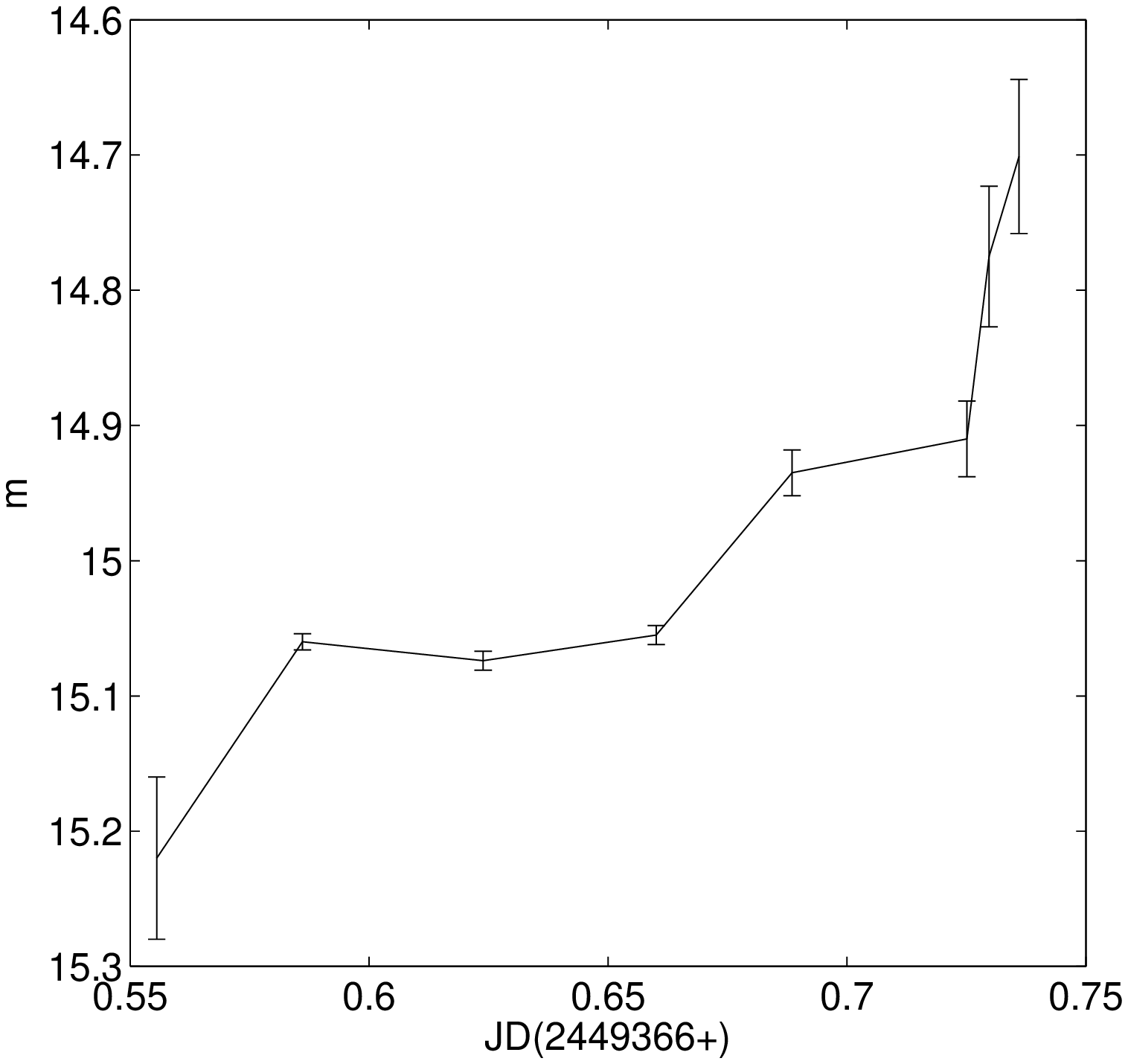}
\caption{Two examples of IDV of OJ287 from OJ-94 data. \label{fig1}}
\end{figure}

\begin{figure}
\begin{center}
\includegraphics[width=8.5cm,height=8cm]{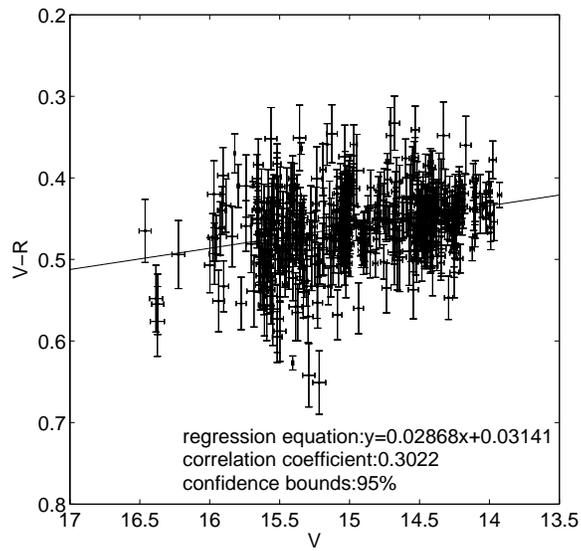}
\caption{Color-magnitude diagram from direct match. The line is the
linear fit to the data. \label{fig2}}
\end{center}
\end{figure}

\begin{figure}
\begin{center}
\includegraphics[width=8.5cm,height=8cm]{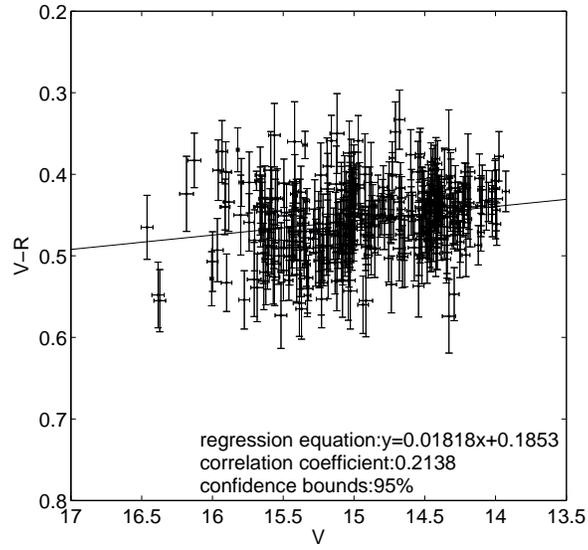}
\caption{Color-magnitude diagram from binning and match. The line is
the linear fit to the data. \label{fig3}}
\end{center}
\end{figure}

\begin{figure}
\begin{center}
\includegraphics[width=14cm,height=8cm]{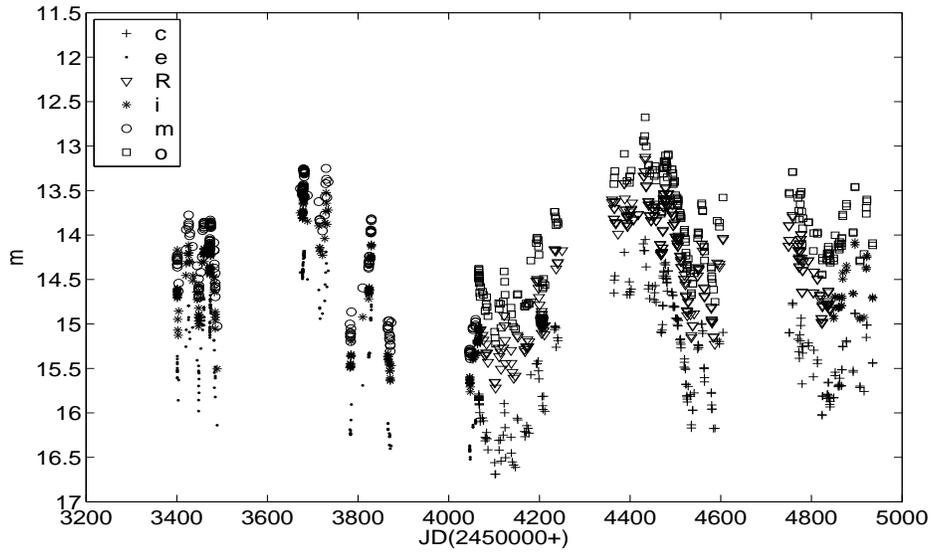}
\caption{Light curves of OJ287 in the BATC $c$, $e$, $R$, $i$, $m$,
and $o$ bands. \label{fig4}}
\end{center}
\end{figure}

\begin{figure}
\begin{center}
\includegraphics[width=8.5cm,height=8cm]{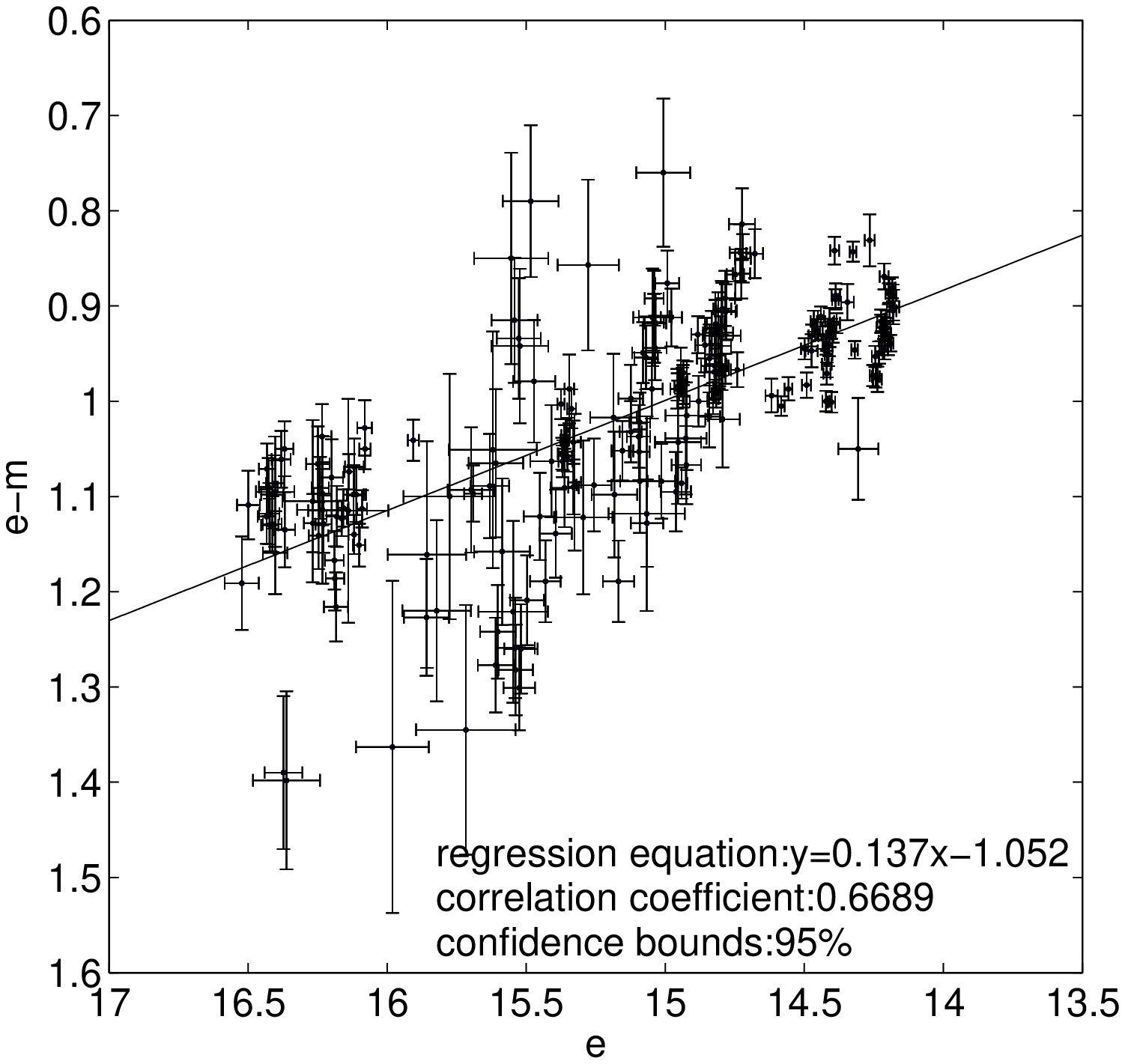}
\caption{Color-magnitude diagram in $e$ and $m$ bands from BATC
data. The line is the linear fit to the data. \label{fig5}}
\end{center}
\end{figure}

\begin{figure}
\begin{center}
\includegraphics[width=8.5cm,height=8cm]{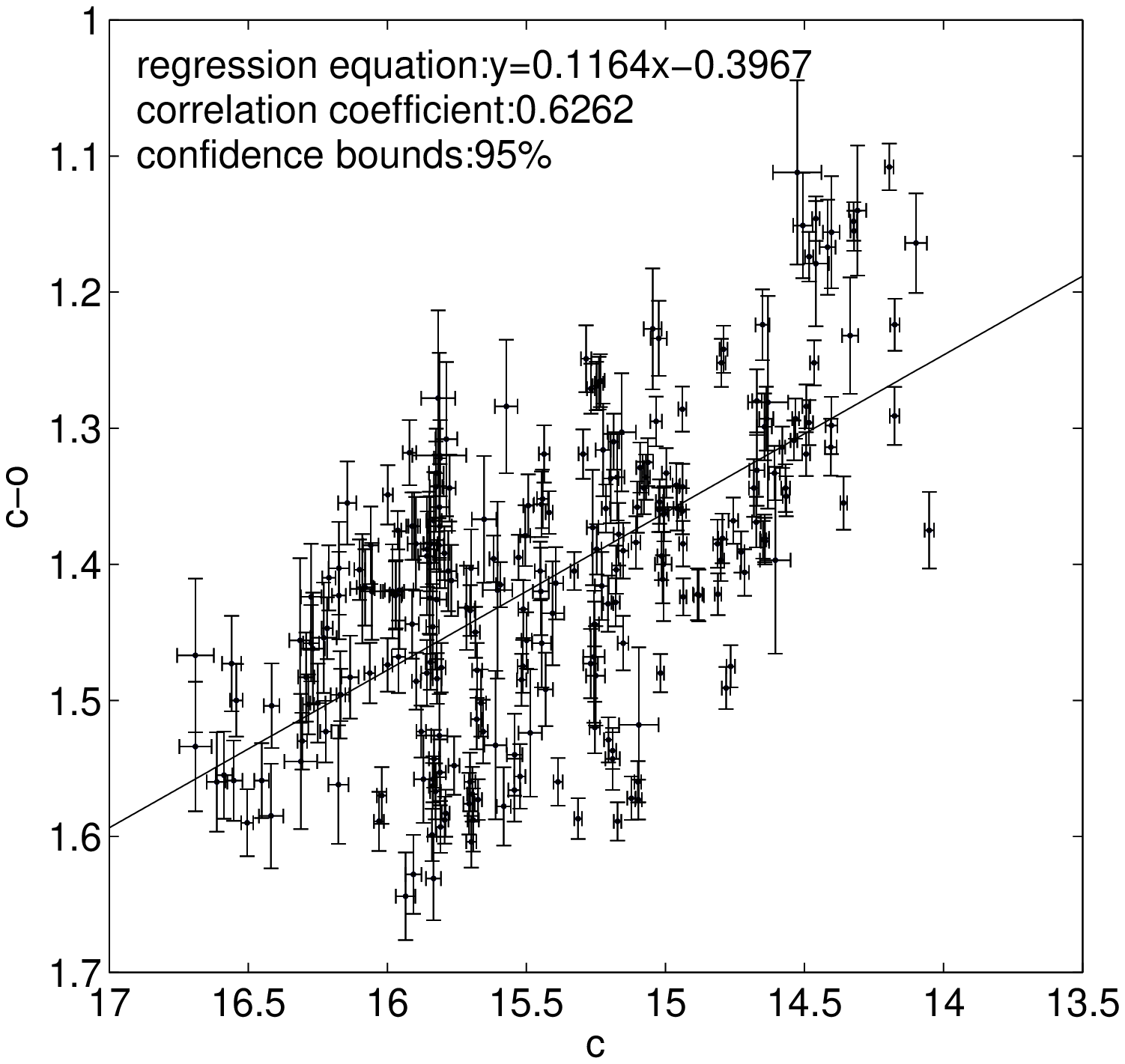}
\caption{Color-magnitude diagram in $c$ and $o$ bands from BATC
data. The line is the linear fit to the data. \label{fig6}}
\end{center}
\end{figure}

\begin{figure}
\includegraphics[width=8cm,height=8cm]{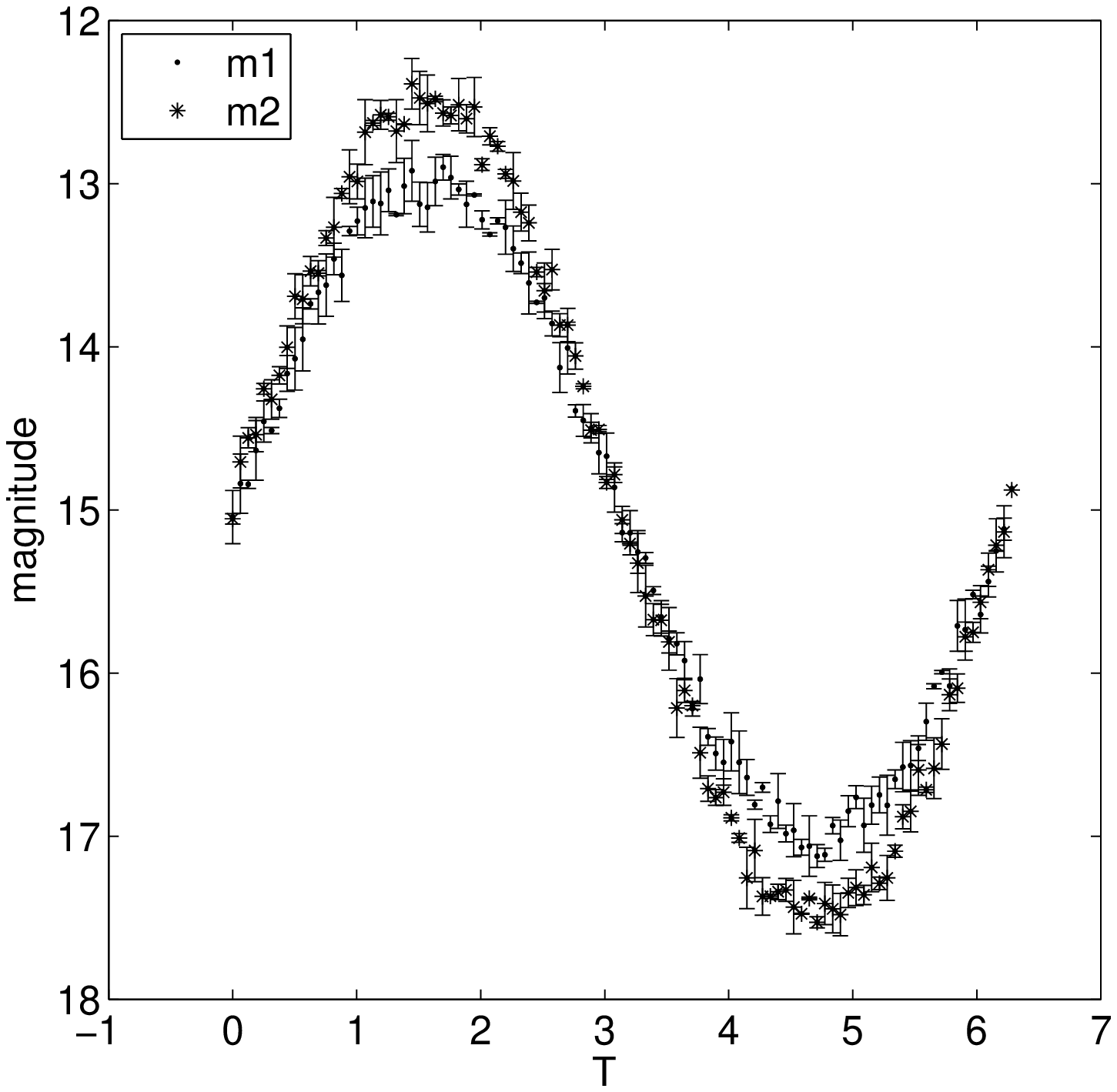}
\includegraphics[width=8cm,height=8cm]{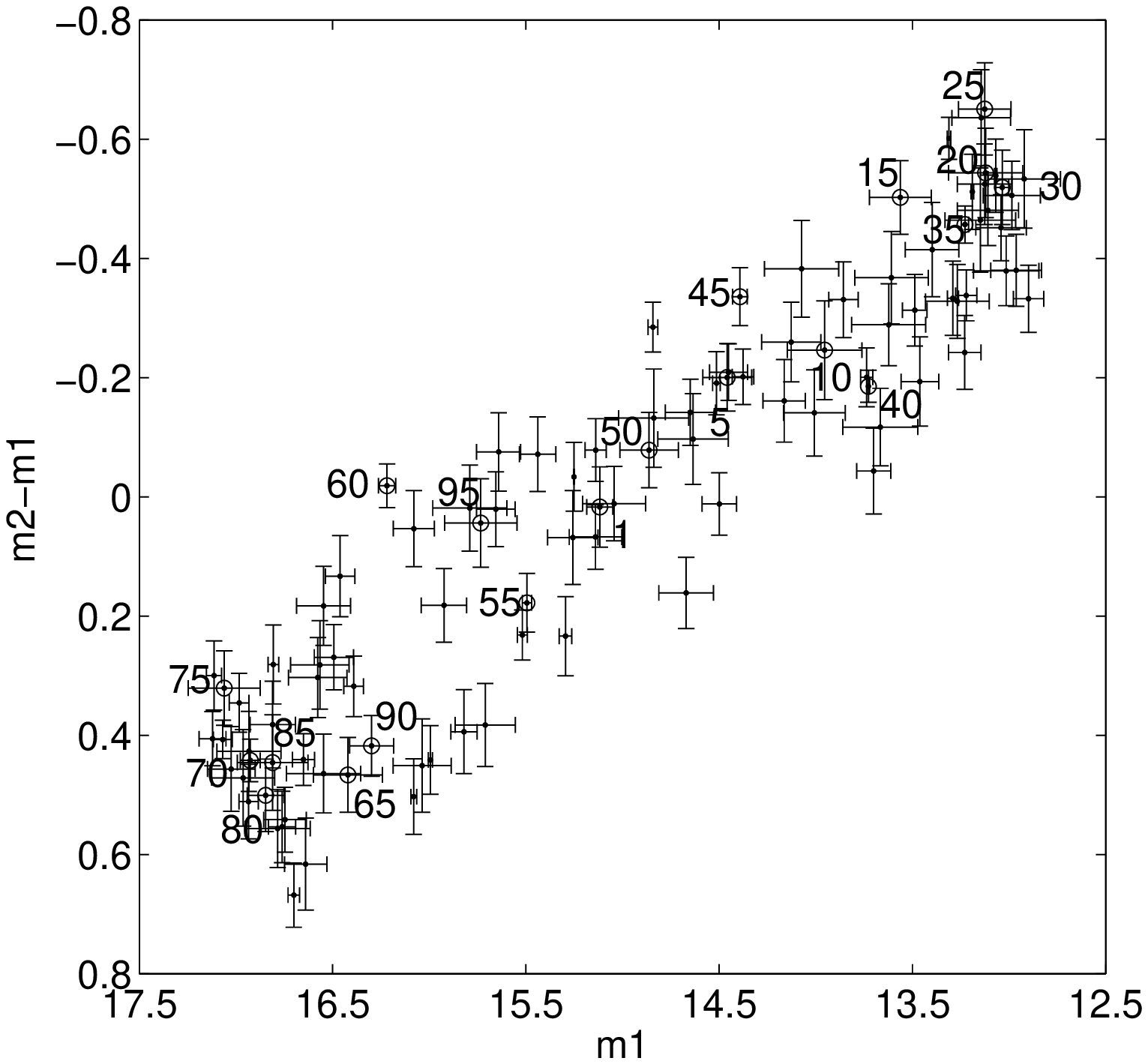}
\caption{Simulated light curves with different amplitudes (left) and
color-magnitude diagram (right). The amplitudes of the two light
curves are 2.0 and 2.5, respectively. \label{fig7}}
\end{figure}

\begin{figure}
\includegraphics[width=8cm,height=8cm]{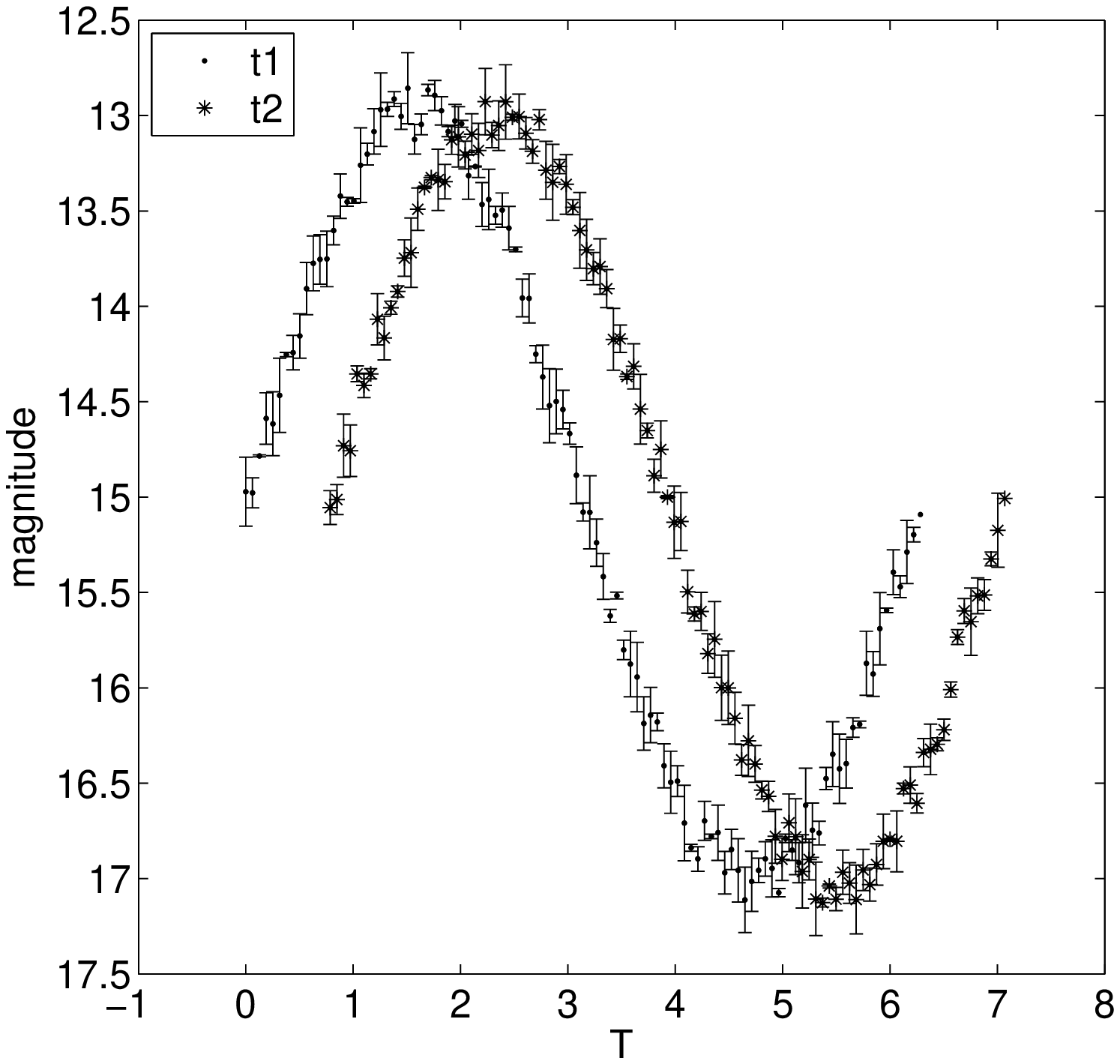}
\includegraphics[width=8cm,height=8cm]{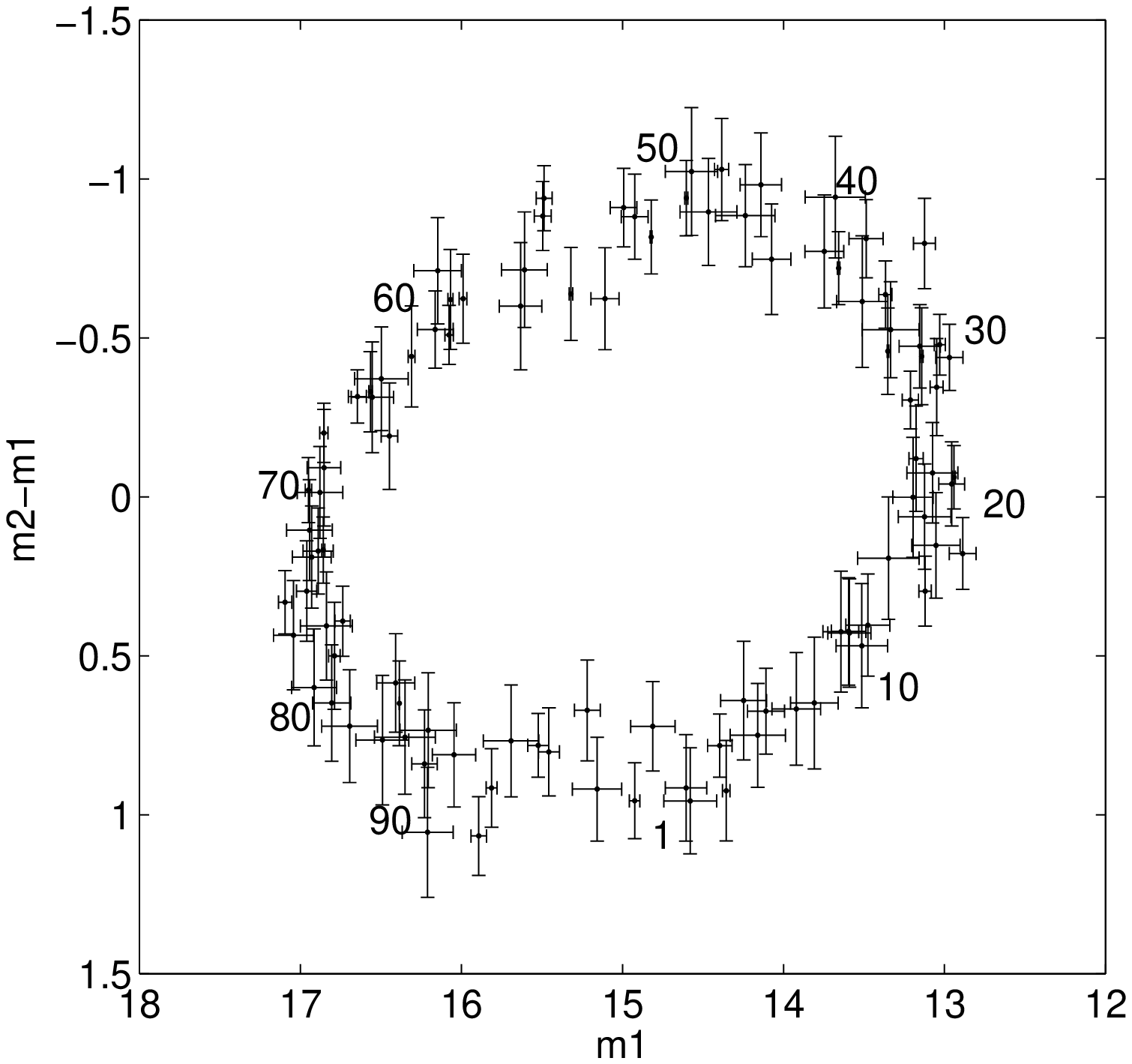}
\caption{Simulated light curves with time delay of 1/8 $\pi$ (left)
and color-magnitude diagram (right). The numbers in the figure
represent time sequence. \label{fig8}}
\end{figure}

\begin{figure}
\includegraphics[width=8cm,height=8cm]{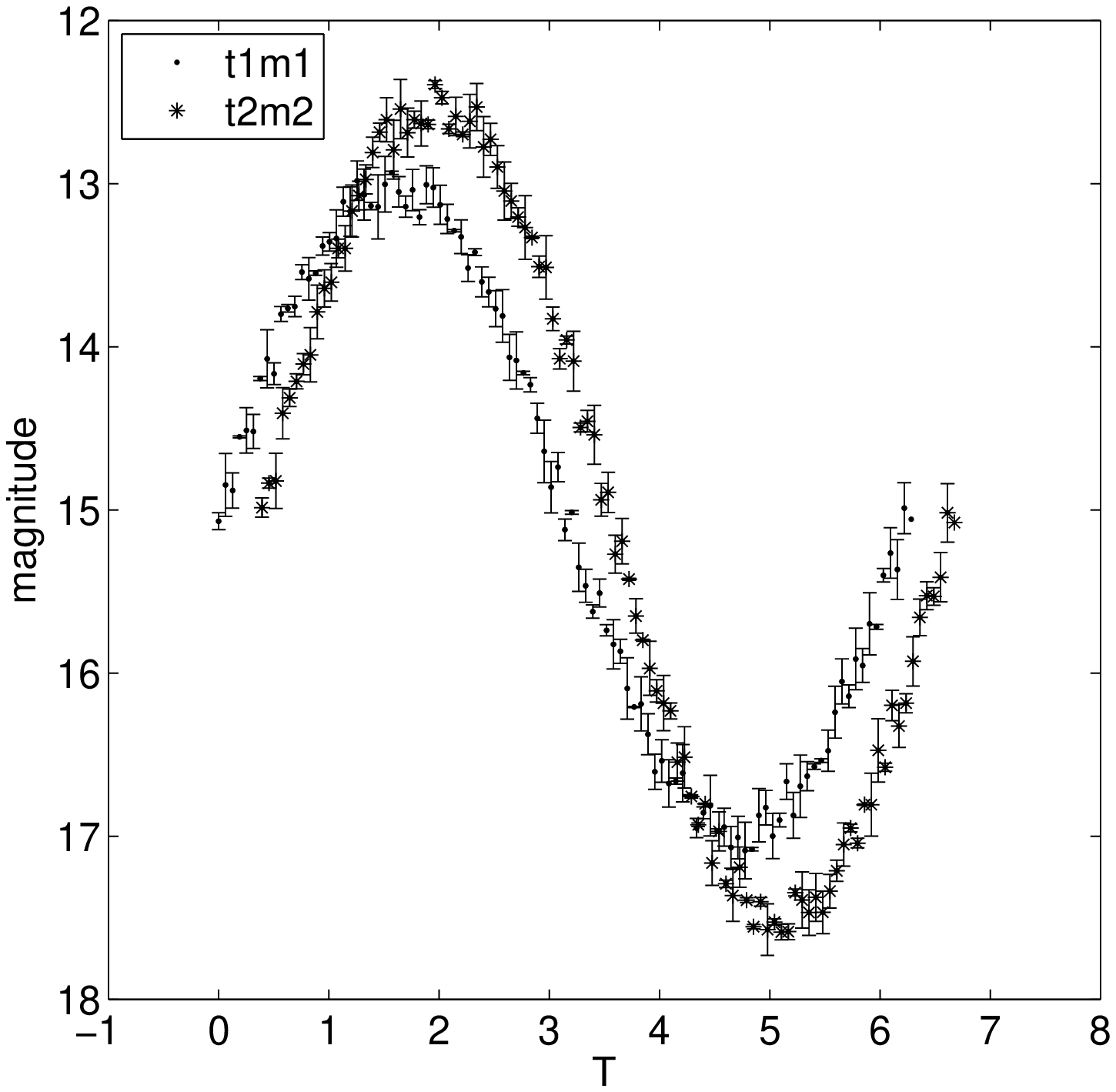}
\includegraphics[width=8cm,height=8cm]{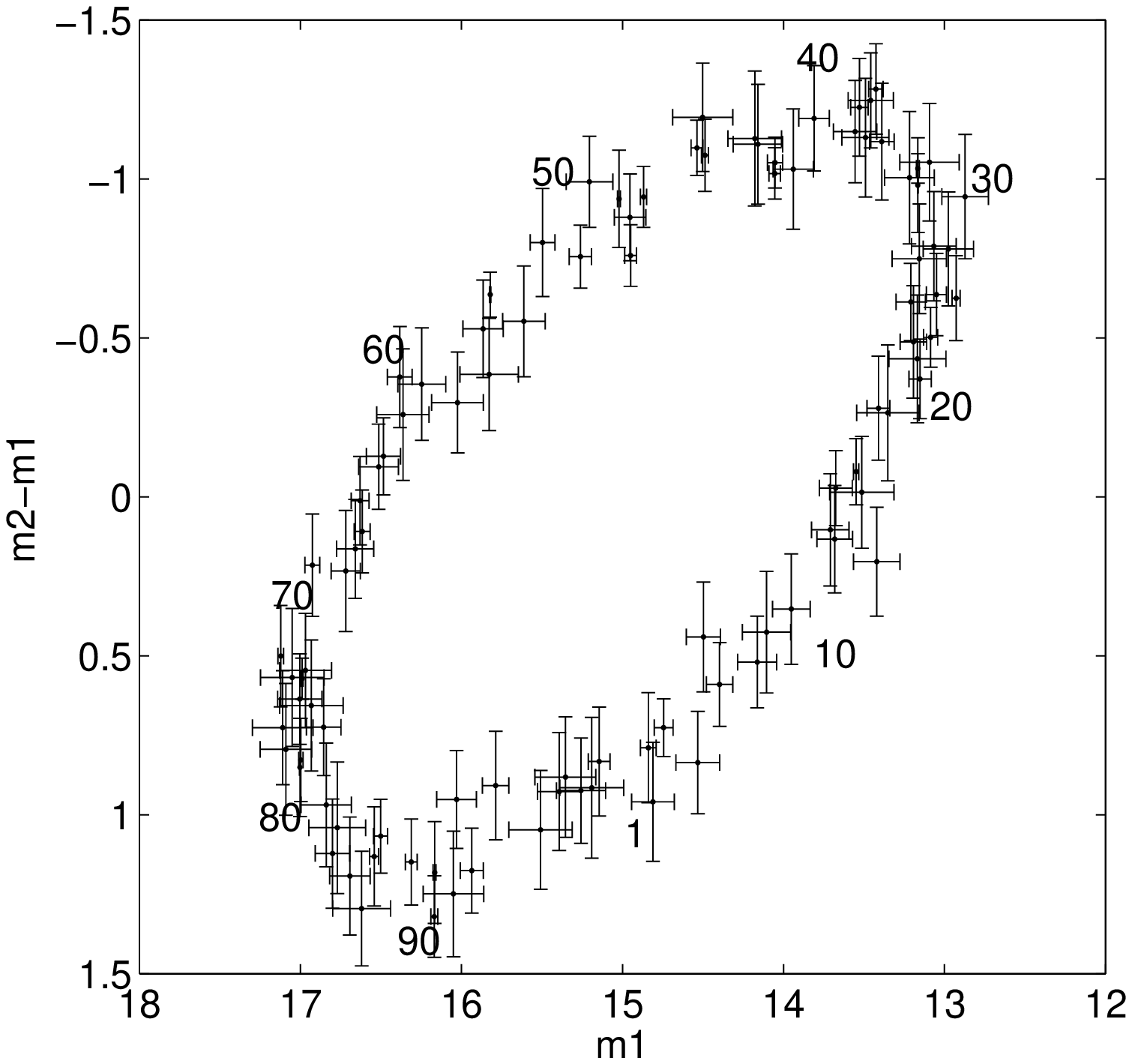}
\caption{Left. Simulated light curves with different amplitudes and
time delay. The amplitudes of the two light curves are 2.0 and 2.5,
respectively, and the time delay is 1/8 $\pi$. Right.
Color-magnitude diagram. The numbers in the figure represent time
sequence. \label{fig9}}
\end{figure}

\clearpage

\begin{deluxetable}{ccccccc}
\tablecaption{BATC magnitudes of the three comparison
stars\label{tbl-1}} \tablewidth{0pt} \tablehead{\colhead{Star ID} &
\colhead{$c$} & \colhead{$e$} & \colhead{$i$} & \colhead{$m$} &
\colhead{$o$}} \startdata
4 &15.198 &14.920 &14.020 &13.650 &13.598 \\
10 &15.082 &15.010 &14.510 &14.280 &14.396 \\
11 &15.479 &15.490 &14.840 &14.590 &14.708 \\
\enddata
\end{deluxetable}

\begin{deluxetable}{ccccccr}
\tablecaption{Data of BATC $c$ band\label{tbl-2}} \tablewidth{0pt}
\tablehead{\colhead{Date(UT)} & \colhead{Time} & \colhead{Julian
Date} & \colhead{Exp} & \colhead{$c$} & \colhead{$c$\_err} &
\colhead{dfmag11}} \startdata
2006 11 26 &18:10:49.0 &2454066.25751 &300 &15.811 &0.043 &0.035 \\
2006 11 26 &18:27:08.0 &2454066.26884 &300 &15.847 &0.035 &$-$0.004 \\
2006 11 26 &18:42:40.0 &2454066.27963 &300 &15.823 &0.035 &$-$0.016 \\
2006 11 26 &18:57:53.0 &2454066.29020 &300 &15.788 &0.039 &$-$0.037 \\
2006 11 26 &19:13:26.0 &2454066.30100 &300 &15.849 &0.037 &0.034 \\
\enddata
\tablecomments{Table \ref{tbl-2} is published in its entirety in the
electronic edition of the {\sl Astronomical Journal}. A portion is
shown here for guidance regarding its form and content.}
\end{deluxetable}

\begin{deluxetable}{ccccccr}
\tablecaption{Data of BATC $e$ band\label{tbl-3}} \tablewidth{0pt}
\tablehead{\colhead{Date(UT)} & \colhead{Time} & \colhead{Julian
Date} & \colhead{Exp} & \colhead{$e$} & \colhead{$e$\_err} &
\colhead{dfmag11}} \startdata
2005 01 29 &14:57:32.0 &2453400.12329 &300 &15.774 &0.249 &0.023 \\
2005 01 29 &15:24:21.0 &2453400.14191 &480 &15.628 &0.236 &$-$0.023\\
2005 01 30 &16:32:48.0 &2453401.18944 &240 &15.549 &0.073 &$-$0.134\\
2005 01 30 &16:44:07.0 &2453401.19730 &240 &15.520 &0.071 &0.024\\
2005 01 30 &16:58:03.0 &2453401.20698 &240 &15.551 &0.066 &0.101\\
\enddata
\tablecomments{Table \ref{tbl-3} is published in its entirety in the
electronic edition of the {\sl Astronomical Journal}. A portion is
shown here for guidance regarding its form and content.}
\end{deluxetable}

\begin{deluxetable}{ccccccr}
\tablecaption{Data of BATC $i$ band\label{tbl-4}} \tablewidth{0pt}
\tablehead{\colhead{Date(UT)} & \colhead{Time} & \colhead{Julian
Date} & \colhead{Exp} & \colhead{$i$} & \colhead{$i$\_err} &
\colhead{dfmag11}} \startdata
2005 01 29 &15:02:48.0 &2453400.12694 &180 &14.683 &0.087 &0.000 \\
2005 01 30 &16:28:41.0 &2453401.18659 &150 &14.632 &0.045 &$-$0.051\\
2005 01 30 &16:48:34.0 &2453401.20039 &150 &14.705 &0.047 &0.025\\
2005 01 30 &17:16:22.0 &2453401.21970 &150 &14.668 &0.046 &$-$0.024\\
2005 01 30 &17:30:36.0 &2453401.22958 &150 &14.631 &0.048 &0.050\\
\enddata
\tablecomments{Table \ref{tbl-4} is published in its entirety in the
electronic edition of the {\sl Astronomical Journal}. A portion is
shown here for guidance regarding its form and content.}
\end{deluxetable}

\begin{deluxetable}{ccccccr}
\tablecaption{Data of BATC $m$ band\label{tbl-5}} \tablewidth{0pt}
\tablehead{\colhead{Date(UT)} & \colhead{Time} & \colhead{Julian
Date} & \colhead{Exp} & \colhead{$m$} & \colhead{$m$\_err} &
\colhead{dfmag11}} \startdata
2005 01 30 &16:38:03.0 &2453401.19309 &240 &14.374 &0.036 &0.011 \\
2005 01 30 &16:52:59.0 &2453401.20346 &240 &14.276 &0.038 &0.023\\
2005 01 30 &17:06:48.0 &2453401.21306 &240 &14.241 &0.037 &0.022\\
2005 01 30 &17:20:50.0 &2453401.22280 &240 &14.265 &0.035 &$-$0.074\\
2005 01 30 &17:35:00.0 &2453401.23264 &240 &14.259 &0.041 &$-$0.018\\
\enddata
\tablecomments{Table \ref{tbl-5} is published in its entirety in the
electronic edition of the {\sl Astronomical Journal}. A portion is
shown here for guidance regarding its form and content.}
\end{deluxetable}

\begin{deluxetable}{ccccccr}
\tablecaption{Data of BATC $o$ band\label{tbl-6}} \tablewidth{0pt}
\tablehead{\colhead{Date(UT)} & \colhead{Time} & \colhead{Julian
Date} & \colhead{Exp} & \colhead{$o$} & \colhead{$o$\_err} &
\colhead{dfmag11}} \startdata
2006 11 26 &18:20:04.0 &2454066.26394 &300 &14.439 &0.048 &0.050 \\
2006 11 26 &18:36:34.0 &2454066.27539 &300 &14.422 &0.046 &0.067\\
2006 11 26 &18:51:29.0 &2454066.28575 &300 &14.397 &0.051 &0.109\\
2006 11 26 &19:07:10.0 &2454066.29664 &300 &14.480 &0.070 &$-$0.006\\
2006 11 26 &19:22:04.0 &2454066.30699 &300 &14.450 &0.064 &0.101\\
\enddata
\tablecomments{Table \ref{tbl-6} is published in its entirety in the
electronic edition of the {\sl Astronomical Journal}. A portion is
shown here for guidance regarding its form and content.}
\end{deluxetable}

\end{document}